\documentstyle[12pt,aps,prd,preprint]{revtex}
\begin{document}
\title{Mode-coupling in rotating gravitational collapse: 
Gravitational and electromagnetic perturbations}
\author{Shahar Hod}
\address{The Racah Institute for Physics, The
Hebrew University, Jerusalem 91904, Israel}
\date{\today}
\maketitle

\begin{abstract}
We consider the late-time evolution 
of {\it gravitational} and electromagnetic 
perturbations in realistic {\it rotating} Kerr spacetimes. 
We give a detailed analysis of the mode-coupling phenomena in rotating
gravitational collapse. A consequence of this phenomena
is that the late-time tail is dominated 
by modes which, in general,
may have an angular distribution different from the original one. 
In addition, we show that different types of fields have {\it different} decaying
rates. This result turns over the traditional belief (which has been widely
accepted during the last three decades) that the late-time tail of 
gravitational collapse is universal.

\end{abstract}

\section{Introduction}\label{Sec1}

The {\it no-hair conjecture}, introduced by Wheeler in the
early 1970s \cite{Whee}, states that the external field of a black
hole relaxes to a  Kerr-Newman field characterized solely 
by the black-hole mass, charge and angular momentum. 

Price \cite{Price} was the first to analyze the mechanism by which the
spacetime outside a (nearly {\it spherical}) star divests itself of
all radiative multipole moments, and leaves behind a Schwarzschild
black hole; it was demonstrated that all radiative perturbations decay
asymptotically as an inverse power of time, the power indices equal 
$2l+3$ (in absolute value), where $l$ is the multipole order of the
perturbation. These inverse power-law
tails are a direct physical 
consequence of the backscattering of waves off the effective curvature
potential at asymptotically far regions \cite{Thorne,Price}. Leaver
\cite{Leaver} demonstrated that the late-time tail can be
associated mathematically with the existence of a branch cut in the Green's function
for the wave propagation problem. 

The analysis of Price has been extended by many authors. 
We shall not attempt to review the numerous works that have been
written addressing the problem of the late-time evolution of
gravitational collapse. For a partial list of references, see e.g.,
\cite{Bicak,Gundlach1,Ching12,HodPir123,Brad1,Brad2,HodPir4,Andersson,Barack1,Hod1,Gundlach2,BurOr}.

The above mentioned analyses were restricted, however, 
to {\it spherically} symmetric backgrounds. 
It is well-known, however, that realistic stellar objects generally rotate about their
axis, and are therefore not spherical. Thus, 
the nature of the physical process of stellar core collapse to
form a black hole is essentially {\it non-}spheric, and an astrophysically
realistic model must take into account the angular momentum of 
the background geometry.

The corresponding problem of wave dynamics in 
realistic {\it rotating} Kerr spacetimes is much more complicated due to
the lack of spherical symmetry. A first progress has
been achieved only recently
\cite{Krivan1,Krivan2,Ori,Barack2,Hod2,BarOr}. 
Evidently, the most interesting situation from a physical point of
view is the dynamics of {\it gravitational} waves in 
{\it rotating} Kerr spacetimes. Recently, we have begun an analytic study of
this fascinating problem \cite{Hod3}. This was done by analyzing the asymptotic late-time
solutions of Teukolsky's master equation \cite{Teukolsky1,Teukolsky2},
which governs the evolution of massless perturbations fields in Kerr spacetimes. In this
paper we give a detailed analysis of the problem. In particular, we
give a full account to the phenomena of mode coupling in rotating
spacetimes (this phenomena has been observed in 
numerical solutions of Teukolsky's equation \cite{Krivan1,Krivan2}) 

The plan of the paper is as follows. In Sec. \ref{Sec2} we give a
short description of the physical system and summarize the main 
analytical results presented in Ref. \cite{Hod3}. In Sec. \ref{Sec3} we discuss
the effects of rotation and the mathematical tools needed for
the physical analysis are derived. 
In Sec. \ref{Sec4} we analyze the active coupling of different 
gravitational and electromagnetic modes 
during a rotating gravitational collapse, with pure initial data. 
In Sec. \ref{Sec5} we consider the late-time evolution of realistic
rotating gravitational collapse, with generic initial data. 
We conclude in Sec. \ref{Sec6} with a summary of our analytical
results and their physical implications.

\section{Review of recent analytical results}\label{Sec2}

The dynamics of massless perturbations outside a realistic rotating
Kerr black hole is governed by Teukolsky's 
master equation \cite{Teukolsky1,Teukolsky2}:

\begin{eqnarray}\label{Eq1}
&& \left[{{(r^{2}+a^{2})^{2}} \over {\Delta}} 
-a^{2}{\rm sin}^{2}\theta \right]
{{\partial ^2 \psi} \over {\partial t^2}} 
+{{4Mar} \over {\Delta}} 
{{\partial ^2 \psi} \over {\partial t \partial \varphi}}
+ \left[{{a^{2}} \over \Delta} -{1 
\over {{\rm sin}^{2} \theta}} \right] {{\partial ^2 \psi} 
\over {\partial \varphi ^2}} \nonumber \\
&& -\Delta^{-s} {\partial \over {\partial r}} 
\left( \Delta^{s+1} {{\partial \psi} 
\over {\partial r}} \right) -{1 \over {{\rm sin}\theta}} 
{{\partial} \over {\partial
    \theta}} \left( {\rm sin} \theta{{\partial \psi} \over {\partial
    \theta}} \right) -2s \left[{{a(r-m)} \over \Delta} 
+{{i{\rm cos} \theta} \over {{\rm sin}^{2} \theta
}} \right] {{\partial \psi} \over {\partial
    \varphi}} \nonumber \\
&& -2s \left[{{M(r^{2}-a^{2})} \over \Delta} -r -ia{\rm cos} 
\theta \right] {{\partial \psi} \over
 {\partial t}} +(s^{2}\cot^{2} \theta -s) \psi =0\ \ ,
\end{eqnarray}
where $M$ and $a$ are the mass and angular-momentum per unit-mass of
the black hole, and $\Delta=r^{2} -2Mr+a^{2}$. (We use gravitational
units in which $G=c=1$). The parameter $s$ is called the spin-weight of the field. For
gravitational perturbations $s= \pm 2$, while for electromagnetic
perturbations $s= \pm 1$. The field quantities $\psi$ which satisfy Teukolsky's equation 
are given in \cite{Teukolsky2}.

Resolving the field in the form

\begin{equation}\label{Eq2}
\psi= \Delta^{-s/2} (r^{2}+a^{2})^{-1/2} 
\sum\limits_{m= -\infty}^{\infty} 
\Psi^{m} e^{im \varphi}\ \ ,
\end{equation}
where $m$ is the azimuthal number, 
one obtains a wave-equation for each value of $m$ (we suppress the index $m$)

\begin{equation}\label{Eq3}
D \Psi \equiv \left [ B_{1} {{\partial ^2} \over {\partial t^2}} +
B_{2} {{\partial} \over
{\partial t}} -
{{\partial ^2} \over {\partial y^2}}+ B_{3} - 
{{\Delta} \over {(r^{2}+a^{2})^2}} {1 \over {{\rm sin}\theta}} 
{{\partial} \over {\partial
    \theta}} \left( {\rm sin} \theta{{\partial} \over {\partial
    \theta}} \right) \right ] \Psi=0\  ,
\end{equation}
where the tortoise radial coordinate $y$ is defined 
by $dy={{r^{2}+a^{2}} \over \Delta} dr$.
The coefficients $B_{i}(r,\theta)$ are given by

\begin{equation}\label{Eq4}
B_{1}(r,\theta)=1-{{\Delta a^{2}{\rm sin}^{2}\theta} \over {(r^{2}+a^{2})^{2}}}\ \ ,
\end{equation}
and

\begin{equation}\label{Eq5}
B_{2}(r,\theta)=\Bigg \{ {{4iMmar} \over \Delta} -2s \left[
  {{M(r^{2}-a^{2})} 
\over \Delta} -r -ia{\rm cos} \theta \right] \Bigg \} {\Delta \over
{(r^2+a^2)^2}}\  .
\end{equation}
[The explicit expression of $B_3(r,\theta)$ is not important for the analysis].

The time-evolution of a wave-field 
described by Eq. (\ref{Eq3}) is given by

\begin{eqnarray}\label{Eq6}
\Psi (z,t) &=& 2\pi \int \int_{0}^{\pi} \Bigg\{ B_{1}(z') 
\Big [ G(z,z';t) \Psi _t(z',0)+G_t(z,z';t) 
\Psi (z',0) \Big] +  \nonumber \\ 
&& B_{2}(z') G(z,z';t) \Psi (z',0) \Bigg\} {\rm sin}\theta' d\theta'dy'\ \ ,
\end{eqnarray}
for $t>0$, where $z$ stands for $(y,\theta)$. The (retarded) Green's 
function $G(z,z';t)$ is defined by  
$DG(z,z';t)=\delta (t) \delta(y-y') {{\delta(\theta - \theta')}/ {2\pi
    {\rm sin}\theta}}$, with $G=0$ for $t<0$. 
We express the Green's function in terms of the the Fourier transform
$\tilde G_{l}(y,y';\omega)$

\begin{equation}\label{Eq7}
G(z,z';t)={1 \over {(2 \pi)^{2}}} \sum\limits_{l=l_{0}}^{\infty} 
\int_{- \infty +ic}^{\infty +ic}
\tilde G_{l}(y,y';\omega)_{s}S_{l}^{m}(\theta,a\omega)_{s}S_{l}^{m}(\theta',a\omega)
e^{-i\omega t} d{\omega}\  ,
\end{equation}
where $c$ is some positive constant and $l_{0}={\rm max}(|m|,|s|)$. 
The functions $_sS_{l}^{m}(\theta,a\omega)$ are the spin-weighted
spheroidal harmonics which are solutions to the angular equation \cite{Teukolsky2}

\begin{eqnarray}\label{Eq8}
&& {1 \over {{\rm sin}\theta}} 
{{\partial} \over {\partial
    \theta}} \left( {\rm sin} \theta{{\partial} \over {\partial
    \theta}} \right) + \nonumber \\
&&  \left(a^{2}{\omega}^{2}{\rm cos}^{2}\theta-{{m^{2}} \over
  {{\rm sin}^{2}\theta}} -2a\omega s{\rm cos} \theta - {{2ms{\rm cos}\theta} \over
  {{\rm sin}^{2}\theta}} - s^{2}\cot^{2}\theta + s +{_{s}A_{l}^{m}} \right) 
{ _{s}S_{l}^{m}}  =0  \  .
\end{eqnarray}
For the $a\omega=0$ case, the eigenfunctions $_{s}S_{l}^{m}(\theta,a\omega)$
reduce to the spin-weighted spherical harmonics
$_{s}Y_{l}^{m}(\theta,\phi)={_{s}S_{l}^{m}(\theta)e^{im\varphi}}$, and the
separation constants $_{s}A_{l}^{m}(a\omega)$ are 
simply $_{s}A_{l}^{m}=(l-s)(l+s+1)$ \cite{Goldberg}. 

The Fourier transform is analytic in the upper half $w$-plane and it satisfies
the equation \cite{Teukolsky2}

\begin{eqnarray}\label{Eq9}
 \tilde D(\omega) \tilde G_l  &\equiv&  \Bigg \{{{d^2} \over {dy^2}} + \left[ {{K^{2} -2is(r-M)K+
    \Delta(4ir\omega s-\lambda)} \over {(r^{2}+a^{2})^{2}}}-H^{2}-{{dH} \over
    {dy}} \right] \Bigg\} \tilde G_{l}(y,y';\omega) \nonumber \\
& =& \delta(y-y')  \  ,
\end{eqnarray}
where $K=(r^{2}+a^{2})\omega-am$, $\lambda=A+a^{2}{\omega}^{2}-2am\omega$, and 
$H=s(r-M)/(r^{2}+a^{2})+r\Delta/(r^{2}+a^{2})^{2}$.

Define two auxiliary functions $\tilde \Psi_1$ and $\tilde \Psi_2$ as 
solutions to the homogeneous equation
$\tilde D(\omega)\tilde \Psi_1$=$\tilde D(\omega)\tilde \Psi_2=0$, with the physical boundary
conditions of purely ingoing waves crossing the event horizon, and
purely outgoing waves at spatial infinity, respectively. In terms of
$\tilde \Psi_1$ and $\tilde \Psi_2$, and henceforth assuming $y'<y$, 
$\tilde G_{l}(y,y';\omega) =-\tilde \Psi_1(y',\omega) \tilde \Psi_2(y,\omega)/W(\omega)$,
where we have used the Wronskian 
relation $W(\omega)=W(\tilde \Psi_1, \tilde \Psi_2)= \tilde \Psi_1 \tilde \Psi_{2,y} - 
\tilde \Psi_2 \tilde \Psi_{1,y}$.

It is well-known that the late-time behaviour of massless perturbations 
fields is determined by the backscattering from asymptotically {\it far}
regions \cite{Thorne,Price}. Thus, the late-time behaviour is dominated by the
{\it low}-frequencies contribution to the Green's function, for only low
frequencies will be backscattered by the small effective curvature 
potential (at $r \gg M$). Therefore, a {\it small}-$\omega$ approximation
[or equivalently, a large-$r$ approximation of Eq. (\ref{Eq9})] is sufficient in order
to study the asymptotic {\it late-time} behaviour of the fields 
\cite{Andersson}. 
With this approximation, the two basic solutions required in order to build 
the Fourier transform are $\tilde \Psi_1 =r^{l +1} e^{i\omega r}
M(l+s+1-2i\omega M ,2l +2, -2i\omega r)$, and 
$\tilde \Psi_2 =r^{l +1} e^{i\omega r} U(l+s+1-2iwM ,2l+2,-2i\omega
r)$, where 
$M(a,b,z)$ and $U(a,b,z)$ are 
the two standard solutions to the confluent hypergeometric equation 
\cite{Abram}. Then

\begin{equation}\label{Eq10}
W(\tilde \Psi_1,\tilde \Psi_2)=i (-1)^{l+1} (2l+1)! (2\omega)^{-(2l
  +1)}/(l+s)!\  .
\end{equation} 

In order to calculate $G(z,z';t)$ using Eq. (\ref{Eq7}), one may 
close the contour of integration into the lower half of the
complex frequency plane. Then, one identifies three distinct contributions to $G(z,z';t)$
\cite{Leaver} : Prompt contribution, quasinormal modes, and tail
contribution. The late-time tail is associated with the existence of a
branch cut (in $\tilde \Psi_2$) in the complex frequency plane \cite{Leaver} (usually
placed along the negative imaginary $\omega-$ axis). A little arithmetic leads to \cite{Hod3}

\begin{eqnarray}\label{Eq11}
\tilde G_l^C(y,y';\omega)&=& \Bigg[{{\tilde \Psi_2(y,\omega e^{2 \pi
      i})} \over {W(\omega e^{2 \pi i})}} -
{{\tilde \Psi_2(y,\omega)} \over {W(\omega)}} \bigg]\tilde \Psi_1(y',\omega)
\nonumber \\
& =&{{(-1)^{l-s} 4 \pi M\omega (l-s)!} \over {(2l+1)! }}
{{\tilde \Psi_1(y,\omega)\tilde \Psi_1(y',\omega)} \over {W(\omega)}}\  .
\end{eqnarray}
Taking cognizance of Eq. (\ref{Eq7}), we obtain

\begin{eqnarray}\label{Eq12}
G^C(z,z';t)& =&  \sum\limits_{l=l_0}^{\infty} 
{{i M (-1)^s 2^{2l+1} (l+s)! (l-s)!} \over {\pi [(2l+1)!]^2}} \nonumber \\
&& \int_{0}^{-i \infty} \tilde \Psi_1(y,\omega) 
\tilde \Psi_1(y',\omega) {_{s}S_{l}(\theta,a\omega)} {_{s}S_{l}(\theta',a\omega)}
{\omega}^{2l+2} e^{-i\omega t} d\omega \  .
\end{eqnarray}

\section{Rotation effects -- the coupling of different modes}\label{Sec3}

The {\it rotational} dragging of reference frames, caused by the rotation
of  the black hole (or star) produces an active {\it coupling} 
between modes of {\it different} $l$ (but the same $m$). 
Mathematically, it is the $\theta$-dependence of 
the spin-weighted spheroidal wave 
functions $_sS_{l}^{m}(\theta,a\omega)$ and of the
coefficients $B_{1}(r,\theta)$ and $B_{2}(r,\theta)$ which is
responsible for the interaction between different modes; {\it no} coupling 
occurs in the non-rotating ($a=0$) case.

The angular equation (\ref{Eq8}) is amenable to a perturbation
treatment for small $aw$ \cite{note1,Hod4}; we write it in the form 
$(L^{0}+L^{1}){_sS_{l}^{m}}=-{_sA_{l}^{m}}{_sS_{l}^{m}}$, where 
$L^{0}(\theta)$ is the $\omega$-{\it independent} part of
Eq. ({\ref{Eq8}), and 

\begin{equation}\label{Eq13}
L^{1}(\theta,a\omega)=(a\omega)^2 {\rm cos}^{2}\theta -2a\omega s {\rm cos} \theta\  ,
\end{equation}
and we use the spin-weighted spherical functions $_sY_{l}^{m}$ as a
representation. They satisfy $L^{0}{_sY_{l}}=-{_sA_{l}^{(0)}}{_sY_{l}}$ 
with $_sA_{l}^{(0)}=(l-s)(l+s+1)$ (we suppress the 
index $m$ on $_sA_l$ and $_sY_l$). 
For small $a\omega$ a standard perturbation theory yields 
(see, for example, \cite{Schiff})

\begin{equation}\label{Eq14}
_sS_{l}(\theta,a\omega)= \sum\limits_{k=l_0}^{\infty} C_{lk}(a\omega)^{|l-k|} {_sY_{k}(\theta)}\  ,
\end{equation}
where, to leading order in $a\omega$, the 
coefficients $C_{lk}(a\omega)$ are $\omega -${\it independent} \cite{Hod2,Hod4}. 
Equation (\ref{Eq14}) implies that the black-hole rotation {\it mixes}
(and ignites) different spin-weighted spherical harmonics.

The coefficients $B_{1}(r,\theta)$ and $B_2(r,\theta)$ appearing in
the time-evolution equation (\ref{Eq6}) depend explicitly on the 
angular variable $\theta$ through the {\it rotation} of the black hole
(no such dependence exist in the non-rotating $a=0$ case). 
Therefore, in order to elucidate the coupling between different modes we should
evaluate the integrals  
$\langle slm|skm \rangle$, $\langle slm|{\rm sin}^2\theta|skm \rangle$, and
$\langle slm|{\rm cos}\theta|skm \rangle$, where 
$\langle slm|F(\theta)|skm \rangle \equiv \int {_sY_l^{m*}}
F(\theta){_sY_k^m} d \Omega$ 
[see Eqs. (\ref{Eq4}) and (\ref{Eq5}) 
for the definition of the $B_i(r,\theta)$ coefficients]. In addition,
the values of the coefficients $C_{lk}$ depend on the integrals 
\cite{Hod2,Hod4} $\langle slm|{\rm cos}^2\theta|skm \rangle$ and
$\langle slm|{\rm cos}\theta|skm \rangle$ [see Eq. (\ref{Eq13}) 
for the definition of the perturbation term $L^{1}(\theta,a\omega)$, which is
responsible for the mixing of modes in rotating backgrounds]. 

The spin-weighted spherical harmonics are related to the rotation
matrix elements of quantum mechanics \cite{CamMor}. Hence, standard
formulae are available for integrating the product of three such
functions (these are given in terms of the Clebsch-Gordan
coefficients \cite{note1,Hod2,Hod4}). In particular, the integrals 
$\langle sl0|{\rm sin}^2\theta|sk0 \rangle$ and $\langle sl0|{\rm cos}^2\theta|sk0 \rangle$ vanish 
unless $l=k,k \pm 2$, while the integral $\langle
sl0|{\rm cos}\theta|sk0 \rangle$ vanishes unless $l=k \pm 1$. 
For non-axially symmetric $(m \neq 0)$ modes, 
$\langle slm|{\rm sin}^2\theta|skm \rangle \neq 0$ for $l=k, k \pm 1, k \pm
2$ (the same holds for the integral $\langle sl0|{\rm cos}^2\theta|sk0 \rangle$), and 
$\langle slm|{\rm cos}\theta|skm \rangle \neq 0$ for $l=k, k \pm 1$ (all
other matrix elements vanish). Note also that the
{\it complex} coefficient $B_2$ couples the real and 
imaginary parts of $\Psi^m$.

We are now in a position to evaluate the late-time evolution of
realistic rotating gravitational collapse. 
We shall consider two kinds of initial data:
Pure initial data, which corresponds to the assumption that the initial angular distribution 
is characterized by a pure spin-weighted spherical harmonic 
function $_sY_{l^*}^m$, and generic initial where the
initial pulse consists of all allowed modes (all spherical harmonics
functions with $l \geq l_0$). 

\section{Pure initial data}\label{Sec4}

\subsection{Asymptotic behaviour at timelike infinity}\label{Sec4A}

As explained, the late-time behaviour of the fields should follow from the
{\it low}-frequency contribution to the Green's function. Actually, it is easy 
to verify that the effective contribution to the integral in 
Eq. (\ref{Eq12}) should come from $|\omega|$=$O({1 /t})$. 
Thus, we may use the $|\omega|r \ll 1$ limit of $\tilde \Psi_1(r,\omega)$ 
in order to obtain the asymptotic
behaviour of the fields at {\it timelike infinity} (where $y,y' \ll t$).
Using Eq. 13.5.5 of \cite{Abram} one 
finds $\tilde \Psi_1(r,\omega) \simeq Ar^{l +1}$. Substituting this 
in Eq. (\ref{Eq12}), and using the representation Eq. (\ref{Eq14})
for the spin-weighted spheroidal wave functions
$_sS_l$, togather with the cited properties of the angular 
integrals (of the form $\langle slm|F(\theta)|skm \rangle$), 
we find that the asymptotic late-time behaviour of 
the $l$ mode (where $l \geq l_0$)
is dominated by the following effective Green's function:

\begin{eqnarray}\label{Eq15}
G^C_l(z,z';t)& =&  \sum\limits_{k=l_0}^{L} 
{{M (-1)^{(l^*+l+2-q-2s)/2} 2^{2k+1} (k+s)! (k-s)! (l^*+l+2-q)!} \over 
{\pi [(2k+1)!]^2}} (yy')^{k+1} \nonumber \\
&& C_{kl}C_{kl^*-q} {_sY_{l}}(\theta) {_sY_{l^*-q}^{*}}(\theta')
a^{l^*+l-2k-q} t^{-(l^*+l+3-q)}\  ,
\end{eqnarray}
where $q={\rm min}(l^*-l_0,2)$. 
Here, $L=l^*-q$ for $l \geq l^*-1$ modes, and $L=l$ for $l \leq l^*-2$
modes. Thus, the late-time behaviour of the gravitational and 
electromagnetic fields at the
asymptotic region of timelike infinity $i_+$ is dominated by the
lowest allowed mode, i.e., by the $l=l_0$ mode. The 
corresponding damping exponent is $-(l^* + l_0+3-q)$.

\subsection{Asymptotic behaviour at future null infinity}\label{Sec4B}

We further consider the behaviour of the fields at the
asymptotic region of future null infinity $scri_+$.
It is easy to verify that for this case
the effective frequencies contributing to the integral in Eq. (\ref{Eq12}) are
of order $O({1 / u})$.
Thus, for $y-y' \ll t \ll 2y-y'$ one may use the
$|\omega|y' \ll 1$ asymptotic limit of $\tilde \Psi_1(y',\omega)$ and 
the $M \ll |\omega|^{-1} \ll y$
($Im \omega < 0$) asymptotic limit of $\tilde \Psi_1(y,\omega)$. Thus, 
$\tilde \Psi_1(y',\omega) \simeq Ay'^{l +1}$, and 
$\tilde \Psi_1(y,\omega) \simeq e^{i\omega y} (2l+1)! e^{-i \pi
  (l+s+1)/2}(2\omega)^{-(l+s+1)}y^{-s}/(l-s)!$, 
where we have used Eqs. 13.5.5 and 13.5.1 of \cite{Abram}, respectively.
Substituting this in Eq. (\ref{Eq12}), and 
using the representation Eq. (\ref{Eq14}) for the spin-weighted 
spheroidal wave functions $_sS_{l}$, togather with the cited properties of the
angular integrals, one finds that the 
behaviour of the $l$ mode (where $l \geq l_0$)
along the asymptotic region of null infinity $scri_+$ 
is dominated by the following effective Green's functions:

\begin{eqnarray}\label{Eq16}
G^C_l(z,z';t)& =&  \sum\limits_{k=l^*-q_1}^{l^*+q_2} 
{{M (-1)^{(l+k-2s+2)/2} 2^{k} (k+s)! (l-s+1)! 
\over {\pi (2k+1)!}}} y'^{k+1}v^{-s} \nonumber \\
&& C_{kl} {_sY_l}(\theta) {_sY^{*}_k}(\theta')
a^{l-k}  u^{-(l-s+2)}\  ,
\end{eqnarray}
for $l \geq l^*-1$ modes, where $q_1={\rm min}(l^*-l_0,2)$ and
$q_2={\rm min}(l-l^*,2)$, and

\begin{eqnarray}\label{Eq17}
G^C_l(z,z';t)& =& {{M (-1)^{(l^*+l-2s)/2} 2^{l} (l+s)!(l^*-s-1)!}
    \over {\pi (2l+1)!}} y'^{l+1}v^{-s} \nonumber \\
&& C_{ll^*-2} {_sY_l}(\theta) {_sY^{*}_{l^*-2}}(\theta')
a^{l^*-l-2} u^{-(l^*-s)}\  ,
\end{eqnarray}
for $l \leq l^*-2$ modes. The dominant modes at null infinity and the
corresponding damping exponents are given in Table \ref{Tab1}. 

\subsection{Asymptotic behaviour along the black-hole outer horizon}
\label{Sec4C}

The asymptotic solution to the homogeneous equation 
$\tilde D(\omega)\tilde \Psi_1(y,\omega)=0$ at the black-hole outer horizon $H_+$
($y \to -\infty$) is \cite{Teukolsky2} 
$\tilde \Psi_1(y,\omega)=C(\omega) \Delta^{-s/2} e^{-i(\omega -m{\omega}_+)y}$, 
where ${\omega}_+=a/(2Mr_+)$ [$r_+=M+(M^2-a^2)^{1/2}$ is
the location of the black-hole outer horizon]. In addition, 
we use $\tilde \Psi_1(y',\omega) \simeq Ay'^{l+1}$. Regularity of the
solution requires $C$ to be an analytic function of $\omega$. We thus
expand $C(\omega)=C_0+C_1{\omega}+\cdots$ for small $\omega$ (as already explained, 
the late-time behaviour of the field is dominated by the {\it low}-frequency
contribution to the Green's function).

Substituting this into Eq. (\ref{Eq12}), and 
using the representation Eq. (\ref{Eq14}) for the spin-weighted 
spheroidal wave functions $_sS_{l}$, we find that the asymptotic 
behaviour of the $l$ mode (where $l \geq l_0$)
along the black-hole outer horizon $H_+$ 
is dominated by the following effective Green's function:

\begin{eqnarray}\label{Eq18}
G^C_l(z,z';t)& =&  \sum\limits_{k=l_0}^{L} 
{_s\Gamma_k} {{M (-1)^{(l^*+l+2-q-2s)/2} 2^{2k+1} (k+s)! 
(k-s)! (l^*+l+2-q)!} \over 
{\pi [(2k+1)!]^2}} \Delta^{-s/2} y'^{k+1} \nonumber \\
&& C_{kl}C_{kl^*-q} {_sY_{l}}(\theta) {_sY_{l^*-q}^{*}}(\theta')
a^{l^*+l-2k-q} e^{imw_+y} v^{-(l^*+l+3-q+b)}\  ,
\end{eqnarray}
where $q, p$ and $L$ are defined as before, $_s\Gamma_k$ are
constants, and $b=0$ {\it generically}, 
except for the unique case $m=0$ with $s>0$, in which $b=1$
\cite{BarOr2}. Hence, the late-time behaviour of the gravitational and
electromagnetic fields along the black-hole outer
horizon is dominated by the
lowest allowed mode, i.e., by the $l=l_0$ mode. The 
corresponding damping exponent is $-(l^* + l_0+3-q+b)$.

\section{Generic initial data}\label{Sec5}

So far we have assumed that the initial pulse is made of {\it pure} data,
characterized by one particular spherical harmonic 
function $_sY_{l^*}^m$. In this section we consider the generic case. 
That is, we assume that the initial pulse consists of 
all the allowed ($l \geq l_0$) modes (see also the most 
recent analysis of Barack \cite{Bar3}). 

The analysis here is very similar to the one presented in
Sec. \ref{Sec4}: Using Eq. (\ref{Eq12}), togather with the appropriate 
asymptotic forms of $\tilde \Psi_1(y,\omega)$ and $\tilde \Psi_1(y',\omega)$ (as given in
Sec. \ref{Sec4} for the various asymptotic regions), 
and the representation Eq. (\ref{Eq14}) for the spheroidal
wave functions, we find that the asymptotic late-time
behaviour of the $l$ mode (where $l \geq l_0$) 
is dominated by the following effective Green's functions:

\begin{equation}\label{Eq19}
G^C_l(z,z';t) =  M F_1 (yy')^{l_0+1} {_sY_{l}}(\theta) {_sY^{*}_{l_0}}(\theta')
a^{l-l_0} t^{-(l+l_0+3)}\  ,
\end{equation}
at {\it timelike infinity} $i_+$, 
where $F_1=F_1(l,l_0,m,s)=(-1)^{(l+l_0+2s+2)/2} 2^{2l_0+1} (l+l_0+2)! (l_0+s)! (l_0-s)!C_{{l_0}l}/\pi [(2l_0+1)!]^2$,

\begin{equation}\label{Eq20}
G^C_l(z,z';t) =  \sum\limits_{k=l_0}^{l} 
M F_2 y'^{k+1}v^{-s} {_sY_l}(\theta) {_sY^{*}_k}(\theta')
a^{l-k}  u^{-(l-s+2)}\  ,
\end{equation}
at future null infinity $scri_+$, 
where $F_2=F_2(l,k,m,s)=(-1)^{(l+k+2s+2)/2} 2^{k} (k+s)!
(l-s+1)!C_{kl}/\pi (2k+1)!$, and

\begin{equation}\label{Eq21}
G^C_l(z,z';t) =  
{_s\Gamma'_l} M F_1 \Delta^{-s/2} y'^{l_0+1}
{_sY_{l}}(\theta) {_sY^{*}_{l_0}}(\theta')
a^{l-l_0} e^{im{\omega}_+y} v^{-(l+l_0+3+b)}\  ,
\end{equation}
at the black-hole outer horizon $H_+$, where $_s\Gamma'_l$ are constants

\section{Summary and physical implications}\label{Sec6}

We have analyzed the dynamics of {\it gravitational} 
(physically, the most interesting case) 
and electromagnetic fields in realistic {\it rotating} black-hole
spacetimes. The main results and their physical implications are as follows:

(1) We have shown that the late-time evolution of 
realistic rotating gravitational collapse is characterized by 
inverse power-law decaying tails at the three asymptotic
regions: timelike infinity $i_{+}$, future null infinity $scri _{+}$, 
and the black-hole outer-horizon $H_{+}$ (where the power-law
behaviour is multiplied by an oscillatory term, caused by the dragging
of reference frames at the event horizon). The relaxation of the
fields is in accordance with the {\it no-hair} conjecture
\cite{Whee}. This work reveals the {\it dynamical} physical
mechanism behind this conjecture in the context of rotating
gravitational collapse.

The dominant modes at asymptotic late-times 
and the values of the corresponding damping exponents are 
summarized in Table \ref{Tab1} (for pure initial data) 
and Table \ref{Tab2} (for generic initial data). 
For reference we also include in Table \ref{Tab3} the results for the scalar field toy
model with pure initial data (the $s=0$ case) \cite{Hod2,BarOr} (the
results for generic initial data coincide with those of gravitational
and electromagnetic perturbations). In these tables, $l$ is the
multipole order of the perturbation, $l_{0}={\rm max}(|m|,|s|)$, and
$l^*$ is the initial mode of the perturbation (for pure initial
data). For the scalar field case ($s=0$), we have $p=0$ if 
$l-|m|$ is even, and $p=1$ otherwise. Note that for pure initial data,
the pulse with $l^*=l_0,l_0+1$ differs from initial data with 
$l_0+2 \leq l^*$. This is caused by the fact that the $l_0$ mode is
not ignited (not coupled) to modes with smaller values of $l$.

The somewhat different character of the scalar field case can be traced back to Eq.  (\ref{Eq5})
for $B_2(r,\theta)$ and Eq. (\ref{Eq13}) for $L^{1}(\theta,a\omega)$; it
turns out that $B_2$ is $\theta$-{\it independent} in the $s=0$ case,
and thus this term cannot couple different modes. To this we should add the
fact that for the scalar field case, $L^{1}(\theta,a\omega)$ is proportional to
$(a\omega)^2$ (the term proportional to $aws$ vanishes), and thus the
coefficients $C_{lk}$ in Eq. (\ref{Eq14}) vanish if $|l-k|$ is odd \cite{Hod2}.

The damping exponents for generic initial data derived in this paper 
agree with those derived most recently by Barack \cite{Bar3} 
using an independent analysis. Note,
however, that Barack's analysis cannot yield the values of the damping
exponents for pure initial data.

(2) The {\it unique} and important feature of {\it rotating}
gravitational collapse (besides the oscillatory
behaviour along the black-hole horizon) 
is the active {\it coupling} of different modes. Physically,
this phenomena is caused by the dragging of reference frames, due to
the black-hole (or star's) rotation (this phenomena is absent in the
non-rotating $a=0$ case). 
As a consequence, the late-time evolution of realistic rotating
gravitational collapse has an angular distribution 
which is generically different from the original angular distribution (in the initial pulse).

(3) We emphasize that the power indices 
at a fixed radius in {\it rotating} Kerr spacetimes 
($l+l_0+3$ for generic initial data and $l^*+l+3-q$
for pure initial data) are generically {\it smaller} than the
corresponding power indices (the well-known $2l+3$)
in {\it spherically} symmetric Schwarzschild 
spacetimes. (For generic initial data 
there is an equality only for the lowest allowed mode $l=l_0$, while
for pure initial data there is an equality only for the $l=l_0$ mode
provided it characterizes the initial pulse). 
This implies a {\it slower} decay of perturbations in rotating Kerr
spacetimes. Stated in a more pictorial
way, a rotating Kerr black hole becomes
`bald' slower than a spherically-symmetric Schwarzschild
black hole. From Eq. (\ref{Eq19}) it is easy to see
that the time scale $t_c$ at which the late-time tail of rotating
gravitational collapse is considerably different from the 
corresponding tail of non
rotating collapse (for $l > l_0$ modes) is $t_c=yy'/a$, where
$y'$ is roughly the average location of the initial pulse.

(4) It has been widely accepted that the late-time tail of
gravitational collapse is {\it universal} in the sense that 
it is {\it independent} of the type of the massless field considered (e.g., scalar,
neutrino, electromagnetic, and gravitational). This belief was based
on {\it spherically} symmetric analyses. Our analysis, however,
turns over this point of view. In particular, the power indices $l+l_0+3$ at a fixed
radius found in our analysis are generically {\it different} from those
obtained in the scalar field toy-model \cite{Hod2,BarOr} $l+|m|+p+3$ (where $p=0$ if
$l-|m|$ is even, and $p=1$ otherwise). Thus, different types of fields have
{\it different} decaying-rates. This is a rather surprising
conclusion, which has been overlooked in the last three decades ! 

\bigskip
\noindent
{\bf ACKNOWLEDGMENTS}
\bigskip

I thank Tsvi Piran for discussions. 
This research was supported by a grant from the Israel Science Foundation.

\begin{table}
\caption{Dominant modes and asymptotic damping exponents for
  gravitational and electromagnetic fields -- pure initial data. 
$l_{0}={\rm max}(|m|,|s|)$, and
$l^*$ is the initial mode of the perturbation.}
\label{Tab1}
\begin{tabular}{llcl}
asymptotic region &$l^*$ & dominant mode(s)& damping exponent\\
\tableline
timelike infinity &$l_0 \leq l^* \leq l_0+1$ & $l_0$ & $-(2l_0+3)$ \\
 &$l_0+2 \leq l^*$ & $l_0$ & $-(l^*+l_0+1)$ \\
null infinity & $l_0 \leq l^* \leq l_0+1$& $l_0$ & $-(l_0-s+2)$ \\
 & $l_0+2 \leq l^*$& $l_0\leq l \leq l^*-2$ & $-(l^*-s)$ \\
outer horizon &$l_0 \leq l^* \leq l_0+1$ & $l_0$ & $-(2l_0+3+b)$ \\
 &$l_0+2 \leq l^*$ & $l_0$ & $-(l^*+l_0+1+b)$ \\
\end{tabular}
\end{table}

\begin{table}
\caption{Dominant modes and asymptotic damping exponents -- generic initial data.}
\label{Tab2}
\begin{tabular}{lcl}
asymptotic region & dominant mode& damping exponent\\
\tableline
timelike infinity &$l_0$ &$-(2l_0+3)$ \\
null infinity & $l_0$ & $-(l_0-s+2)$ \\
outer horizon &$l_0$ & $-(2l_0+3+b)$ \\
\end{tabular}
\end{table}

\begin{table}
\caption{Dominant modes and asymptotic damping exponents for scalar
  fields -- pure initial data. $p=0$ if 
$l-|m|$ is even, and $p=1$ otherwise.}
\label{Tab3}
\begin{tabular}{llcl}
asymptotic region &$l^*$ & dominant mode(s)& damping exponent\\
\tableline
timelike infinity &$l_0 \leq l^* \leq l_0+1$ & $l^*$ & $-(2l^*+3)$ \\
 &$l_0+2 \leq l^*$ & $l_0+p$ & $-(l^*+l_0+p+1)$ \\
null infinity & $l_0 \leq l^* \leq l_0+1$& $l^*$ & $-(l^*+2)$ \\
 & $l_0+2 \leq l^*$& $l_0+p\leq l \leq l^*-2$ & $-l^*$ \\
outer horizon &$l_0 \leq l^* \leq l_0+1$ & $l^*$ & $-(2l^*+3)$ \\
 &$l_0+2 \leq l^*$ & $l_0+p$ & $-(l^*+l_0+p+1)$ \\
\end{tabular}
\end{table}


\begin{thebibliography}{99}

\bibitem{Whee}R. Ruffini and J. A. Wheeler, Physics Today {\bf 24}, 30
  (1971); C. W. Misner, K. S. Thorne and J. A. Wheeler, Gravitation
  (Freeman, San Francisco 1973).

\bibitem{Price} R. H. Price, Phys. Rev. D {\bf 5}, 2419 (1972).

\bibitem{Thorne} K. S. Thorne, p. 231 in Magic without magic: John Archibald
Wheeler Ed: J.Klauder (W.H. Freeman, San Francisco 1972).

\bibitem{Leaver} E. W. Leaver, Phys. Rev. D {\bf 34}, 384 (1986).

\bibitem{Bicak}J. Bi\v{c}\'{a}k, Gen. Relativ. Gravitation {\bf 3}, 331 (1972).

\bibitem{Gundlach1} C. Gundlach, R.H. Price, and J. Pullin,
  Phys. Rev. D {\bf 49}, 883 (1994).

\bibitem{Ching12} E. S. C. Ching, P. T. Leung, W. M. Suen, and K. Young,
  Phys. Rev. D {\bf 52}, 2118 (1995); Phys. Rev. Lett. {\bf 74}, 2414 (1995).

\bibitem{HodPir123} S. Hod and T. Piran, Phys. Rev. D {\bf 58}, 024017
  (1998); Phys. Rev. D {\bf 58}, 024018 (1998); Phys. Rev. D {\bf 58}, 024019 (1998).

\bibitem{Brad1} P. R. Brady, C. M. Chambers and W. Krivan,
  Phys. Rev. D {\bf 55}, 7538 (1997).

\bibitem{Brad2} P. R. Brady, C. M. Chambers, W. G. Laarakkers 
and E. Poisson, Phys. Rev. D {\bf 60}, 064003 (1999).

\bibitem{HodPir4} S. Hod and T. Piran, Phys. Rev. D {\bf 58}, 044018 (1998).

\bibitem{Andersson} N. Andersson, Phys. Rev. D {\bf 55}, 468 (1997).

\bibitem{Barack1} L. Barack, Phys. Rev. D {\bf 59}, 044017 (1999).

\bibitem{Hod1} S. Hod, Phys. Rev. D {\bf 60}, 104053 (1999).

\bibitem{Gundlach2} C. Gundlach, R.H. Price, and J. Pullin,
  Phys. Rev. D {\bf 49}, 890 (1994).

\bibitem{BurOr} L. M. Burko and A. Ori, Phys. Rev. D {\bf 56}, 7820
  (1997).

\bibitem{Krivan1} W. Krivan, P. Laguna and P. Papadopoulos,
  Phys. Rev. D {\bf 54}, 4728 (1996).

\bibitem{Krivan2} W. Krivan, P. Laguna and P. Papadopoulos and N. Andersson,
  Phys. Rev. D {\bf 56}, 3395 (1997).

\bibitem{Ori} A. Ori, Gen. Rel. Grav. {\bf 29}, Number 7, 881 (1997).
 
\bibitem{Barack2} L. Barack, in Internal structure of black holes and
  spacetime singularities, Volume XIII of the Israel Physical Society,
  Edited by L. M. Burko and A. Ori (Institute of Physics, Bristol, 1997).

\bibitem{Hod2} S. Hod, e-print gr-qc/9902072, Phys. Rev. D. {\bf 61},
  024033 (2000).

\bibitem{BarOr} L. Barack and A. Ori, e-print gr-qc/9902082,
  Phys. Rev. Lett. {\bf 82}, 4388 (1999).

\bibitem{Hod3} S. Hod, Phys. Rev. D {\bf 58}, 104022 (1998).

\bibitem{Teukolsky1} S. A. Teukolsky, Phys. Rev. Lett. {\bf 29}, 
1114 (1972).

\bibitem{Teukolsky2} S. A. Teukolsky, Astrophys. J. {\bf 185}, 635 (1973).

\bibitem{Goldberg} J. N. Goldberg, A. J. Macfarlane, E. T. Newman,
  F. Rohrlich and E. C. G. Sudarshan, J. Math. Phys. {\bf 8}, 2155
  (1967).

\bibitem{Abram} M. Abramowitz and I.A. Stegun, Handbook of mathematical
functions (Dover Publications, New York 1970).

\bibitem{note1} Many of the technical details here are analogous to
  those discussed in \cite{Hod2} for the scalar field case. 
We therefore give here only the details which are unique to the $s \neq 0$
  case. Full details can be found in \cite{Hod4}.

\bibitem{Hod4} S. Hod, e-print gr-qc/9902073 v1.

\bibitem{Schiff} L. Schiff, Quantum Mechanics, 3rd edition 
(McGraw-Hill, New York, 1968).

\bibitem{CamMor} W. B. Campbell and T. Morgan, Physica {\bf 53}, 264
  (1971).

\bibitem{BarOr2} L. Barack and A. Ori have recently shown,
  Phys. Rev. D {\bf 60}, 124005 (1999), that $C_0$ vanishes in the particular
  case $am=0$ with $s>0$. Thus, 
$b=1$ in this non-generic case, whereas $b=0$ in all other cases.

\bibitem{Bar3} L. Barack, Phys. Rev. D {\bf 61}, 024026 (2000).

\end{thebibliography}
\end{document}